\documentclass[pre]{revtex4}
\usepackage{amsmath}
\usepackage{graphicx}
\begin{document}
\title{Flip motion of solitary wave in an Ising-type Vicsek model}
\author{Hidetsugu Sakaguchi and Kazuya Ishibashi}
\affiliation{Department of Applied Science for Electronics and Materials,
Interdisciplinary Graduate School of Engineering Sciences, Kyushu
University, Kasuga, Fukuoka 816-8580, Japan}
\begin{abstract}
An Ising-type Vicsek model is proposed for collective motion and sudden direction change in a population of self-propelled particles. Particles move on a linear lattice with velocity $+1$ or $-1$ in the one-dimensional model.  The probability of the velocity of a particle at the next step is determined by the number difference of the right- and left- moving particles at the present lattice site and its nearest-neighboring sites. A solitary wave appears also in our model similarly to previous models. In some parameter range, the moving direction of the solitary wave sometimes changes rather suddenly, which is like the sudden change of moving direction of a flock of birds. We study the average reversal time of traveling direction numerically and compare the results with a mean-field theory. The one-dimensional model is generalized to a two-dimensional model. Flip motion of a bandlike soliton is observed in the two-dimensional model.  
\end{abstract}
\maketitle
\section{Introduction}
Collective motion of self-propelled particles has been intensively studied as active matters since Vicsek et al. proposed a simple model for a large population of self-propelled particles~\cite{Vicsek,Vicsek2, Marchetti}. 
Complex and cooperative motions have been experimentally observed in various biological and nonbiological systems~\cite{Sokolov, Scaller, Katz,Thutupalli,Sumino}.
The collective motion appears via an order-disorder transition in the Vicsek model~\cite{gre}.  In the disordered state, the directions of self-propelled particles are randomly distributed. In the ordered state, a certain average direction appears and the self-propelled particles move collectively. The spatial distribution is uniform in the ordered state. However, there is another nonuniform state called a solitary wave state. In the solitary wave state, a highly concentrated region of self-propelled particles propagate like a one-dimensional solitary wave~\cite{Chate,Bertin,Mishra,Gopinath, Bricard, Ihle}. The solitary wave state was first found in direct numerical simulations of the Vicsek model. There are several discussions about the formation mechanism of the solitary wave state, however, it is not clearly understood yet~\cite{Caussin}. In previous papers, we showed that the solitary wave state can appear in the nonlinear Kramers equation~\cite{Sakaguchi} for the probability distribution of the position and velocity of self-propelled particles. Furthermore, we proposed a simple model of one-dimensional solitary wave state, in which the linear instability of the uniform disordered state leads to the formation of a solitary wave state~\cite{Sakaguchi2}.  In these models, the solitary waves move steadily in the right or left direction.  In the original two-dimensional Vicsek model, self-propelled particles can move in any direction. Solon and Tailleur proposed the active Ising model, in which self-propelled particles undergo a diffusion motion biased in the right or left direction and the updating is randomly performed like the Monte-Carlo simulation in the Ising model~\cite{Solon}. In their two-dimensional model, there is a steadily-moving phase-separated state in a parameter range between the disordered and ordered states. They showed a coarsening process to the phase separated state, which implies that the steadily moving localized state is finally attained in the parameter range. However, there are only two states of disordered and flipping cluster states due to strong fluctuations in one dimension~\cite{Solon, Solon2,Solon3}. 

On the other hand, we often observe the moving direction changes collectively in the school of fish and flocks of birds. For example, the flock of starling changes the moving direction rather suddenly.  In this paper, we propose a new simple model between the original Vicsek model and the active Ising model to understand the collective change of moving direction. The self-propelled particles move on a one-directional lattice with velocity $\pm 1$. Our model has the discrete symmetry of the right and left and particles move only on a lattice, which is similar to the active Ising model. The moving direction of a particle changes randomly with probability determined by the number of the right and left moving particles on the site and its nearest-neighbor sites.  We call this model the Ising-type Vicsek model. In our model, the random walk or the diffusion motion is not assumed, the probability of the velocity is determined by the interaction with particles in the surrounding sites, and the updating is performed simultaneously for all particles, which is similar to the original Vicsek model. 
In our model, flip motions of solitary waves occur more clearly, although similar change of moving direction was observed in the flipping cluster state of the one-dimensional active Ising model.  That is, a right-traveling solitary wave is replaced by a left-traveling solitary wave in a very short time. We show some numerical results and discuss the mechanism. The formation of a sharp solitary wave in a very short time in our model is suggestive for the general formation mechanism of solitary waves in a large population of self-propelled particles. The one-dimensional model can be generalized to a two-dimensional model. Flip motions are observed even in the two-dimensional model, which is different from the model of Solon and Tailleur.         
\section{Ising-type Vicsek model}
In the one-dimensional Ising-type Vicsek model, each particle moves on a linear lattice with velocity $\pm 1$. The velocity of the $i$th particle at discrete time $t$ is expressed as $v_{i,t}$, and the position $x_{i,t+1}$ of the $i$th particle at $t+1$ is determined by a simple rule: $x_{i,t+1}=x_{i,t}+v_{i,t+1}$. The total number of particles is expressed as $N$. The lattice size is $L$ and periodic boundary conditions are assumed. 
  
  The number of right (left) moving particles on site $j$ at a discrete time $t$ is denoted as $n_{j,t}^+$ ($n_{j,t}^-)$. The particle number at the site $j$ is denoted as $n_{j,t}=n_{j,t}^{+}+n_{j,t}^{-}$. The number difference $m_{j,t}=n_{j,t}^{+}-n_{j,t}^{-}$ between the right and left moving particles at the site expresses a local order parameter corresponding to magnetization. In the active Ising model, each particle has a spin variable $s_i$, and the particle moves at rate $D(1+s\epsilon)$ and $D(1-s\epsilon)$ to its right and left neighboring site. The spin $s$ is assumed to flip to $-s$ at rate $\exp(-sm_{j,t}/Tn_{j,t})$, where $T$ is a parameter corresponding to the temperature. Both the particle position and the spin variable change randomly, and the flipping rate is determined by the  ratio $m_{j,t}/n_{j,t}$ of the local order parameter over the local particle number. In our model, only the velocity $v_{i,t}$ is a stochastic variable and the position at the next time step $t+1$ is determined by the velocity $v_{i,t+1}$. The particle velocity $v_{i,t}=1$ or -1 is randomly chosen with the probability determined by the interaction with particles in the same site and the neighboring sites. For a local sum of order parameter, $\tilde{m}_{j,t}$ is defined as $\tilde{m}_{j,t}=m_{j,t}+q(m_{j-1,t}+m_{j+1,t})$, where $q$ is a parameter for the strength of the nearest-neighbor coupling. For the sake of simplicity, only numerical results of $q=1/2$ are shown in this paper, although similar results were obtained for $0<q\le 1$. Using $\tilde{m}_{j,t}$, the probability that the $i$th particle on the $j$th site takes $v_{i,t+1}=\pm 1$ at the next time step $t+1$ is assumed to be  
\begin{equation}
p_{j,t+1}^{\pm}=\frac{e^{\pm g\tanh \beta \tilde{m}_{j,t}}}{e^{g\tanh\beta \tilde{m}_{j,t}}+e^{-g\tanh \beta \tilde{m}_{j,t}}},
\end{equation}
where $g>0$ and $\beta>0$ are control parameters. The function $\tanh\beta \tilde{m}_{j,t}$ is introduced simply to connect continuously  two limiting cases explained below. If the number of right-moving particles is larger than that of left-moving particles, then $\tilde{m}_{j,t}>0$ and the probability of $v_{i,t+1}=1$ is larger than that of $v_{i,t+1}=-1$ at $t+1$. This probability represents cooperative interaction similar to ferromagnetic interaction in the Ising model.  The moving directions of particles tend to align owing to this ferromagnetic interaction. By a kind of mean-field approximation, the average numbers $n_{j+1,t+1}^{+}$  and $n_{j-1,t+1}^{-}$ of right- and left- moving particles at time $t+1$ are expressed as 
\begin{eqnarray}
n_{j+1,t+1}^{+}&=&\frac{e^{G_{j,t}}}{e^{G_{j,t}}+e^{-G_{j,t}}}n_{j,t},\nonumber\\
n_{j-1,t+1}^{-}&=&\frac{e^{-G_{j,t}}}{e^{G_{j,t}}+e^{-G_{j,t}}}n_{j,t},
\end{eqnarray}
where $G_{j,t}=g\tanh\beta \tilde{m}_{j,t}=g\tanh[\beta\{n_{j,t}^{+}-n_{j,t}^{-}+q(n_{j-1,t}^{+}-n_{j-1,t}^{-}+n_{j+1,t}^{+}-n_{j+1,t}^{-})\}]$. 

If $\beta$ is sufficiently small, $g\tanh\beta \tilde{m}_{j,t}\sim g\beta \tilde{m}_{j,t}$, and the probability of $v_{i,t+1}=\pm 1$ is determined by the probability   
\begin{equation}
p_{j,t+1}^{\pm}=\frac{e^{\pm g\beta \tilde{m}_{j,t}}}{e^{g\beta \tilde{m}_{j,t}}+e^{-g\beta \tilde{m}_{j,t}}}.
\end{equation}
This probability is similar to that in the equilibrium system, where $g\tilde{m}_{j,t}$ expresses the effective magnetic field by the mutual interaction,  and $\beta$ can be interpreted as the inverse temperature. The term $g\beta\tilde{m}_{j,t}$ in our model corresponds to $m_{j,t}/(T n_{j,t})$ in the active Ising model by Solon and Tailleur. The interaction becomes stronger if the particles gather together, because the division by $n_{j,t}$ is not assumed in our model. On the other hand, we mainly study the simple case of $\beta=\infty$ in the next sections. In this limiting case, $\tanh \beta \tilde{m}_{j,t}$ takes $1$ or -1 depending only on the sign of $\tilde{m}_{j,t}$, and the probability of $v_{i,t+1}=\pm 1$ is given by 
\begin{eqnarray}
p_{j,t+1}^{\pm}&=&\frac{e^{\pm g}}{e^{g}+e^{-g}},\;\;\; {\rm for}\;\; \tilde{m}_{j,t}>0,\nonumber \\
&=&\frac{e^{\mp g}}{e^{g}+e^{-g}},\;\;\; {\rm for}\;\; \tilde{m}_{j,t}<0,
\end{eqnarray}
and $p_{j,t+1}^{\pm}=1/2$ for $\tilde{m}_{j,t}=0$.  
The probability of the velocity at the next time step is determined by the majority rule, that is, particles tend to go along with the majority. However, there is a finite probability of taking the opposite velocity to the majority even for very large $\tilde{m}_{j,t}$ in this model. The function $\tanh\beta \tilde{m}_{j,t}$ has the effect to saturate the interaction strength when $\tilde{m}_{j,t}$ is large.  In this simple case, the local order parameter $m_{j,t+1}$ is 0 in the disordered state, and $m_{j,t+1}>0$ satisfies   
\begin{equation}
m_{j,t+1}=n_{j,t+1}^{+}-n_{j,t+1}^{-}=\frac{N}{L}{\rm tanh}g
\end{equation}
in the spatially-uniform ordered state. 

First, we study the linear stability of the disordered state in the model with the probability Eq.~(1) using the mean-field approximation. 
For sufficiently small $g\beta$, the disordered state is stable, in which the system is spatially uniform and the long-time average of $v_i$ is zero or $\langle v_i \rangle=0$. As $g\beta$ increases, a kind of order-disorder phase transition occurs owing to the cooperative interaction. The average numbers $n_{j+1,t+1}^{+}$  and $n_{j-1,t+1}^{-}$ of right- and left- moving particles obey Eq.~(2) in the mean field approximation. In the disordered state, $n_{j,t}^{+}=n_{j,t}^{-}=N/(2L)$. A linear stability around this uniform state can be investigated by assuming $n_{j,t}^{+}=N/(2L)+\delta n_{k_m,t}^{+}e^{ik_mj}$ and $n_{j,t}^{-}=N/(2L)+\delta n_{k_m,t}^{-}e^{ik_mj}$ where $k_m=2\pi m/L$. The perturbation amplitudes $\delta n_{k_m,t+1}^{+}$ and $\delta n_{k_m,t+1}^{-}$ obey
\begin{eqnarray}
\delta n_{k_m,t+1}^{+}&=& e^{-ik_m}\{1/2+(g\beta N)/(2L)(1+2q\cos k_m)\}\delta n_{k_m,t}^{+}+e^{-ik_m}\{1/2-(g\beta N)/(2L)(1+2q\cos k_m)\}\delta n_{k_m,t}^{-},\nonumber\\
\delta n_{k_m,t+1}^{-}&=& e^{ik_m}\{1/2-(g\beta N)/(2L)(1+2q\cos k_m)\}\delta n_{k_m,t}^{+}+e^{ik_m}\{1/2+(g\beta N)/(2L)(1+2q\cos k_m)\}\delta n_{k_m,t}^{-}.
\end{eqnarray}
The linear growth rate $\lambda_{k_m}$ for the perturbation obeys
\begin{equation}
\lambda_{k_m}^2-\{1+(g\beta N/L)(1+2q\cos k_m)\}\cos k_m\lambda_{k_m}+(g\beta N/L)(1+2q\cos k_m)=0.
\end{equation}
For $k_m=0$, there are two solutions $\lambda_{k_m}=1$ and $\lambda_{k_m}=g\beta N(1+2q)/L$. 
For $k_m=\pi$, there are two solutions $\lambda_{k_m}=-1$ and $\lambda_{k_m}=-g\beta N(1-2q)/L$. If $g\beta>L/\{N(1+2q)\}=L/(2N)$ at $q=1/2$, one eigenvalue $\lambda_{k_m}$ satisfies $\lambda_{k_m}>1$ for $k_m=0$ and the disordered state becomes unstable.

Next, we show numerical results for $N=200$ and $L=500$ for the general model of Eq.~(1). 
Initially, each particle is randomly set on the lattice, and $v_i=1$ or -1  is randomly chosen. Figure 1(a) shows the time evolution of $n_{j,t}^{+}$ at $g=3.7$ and $\beta=0.05$. Although the uniform state becomes linearly unstable only for $g>g_c=L/(2N\beta)=25$, a solitary wave appears even at $g=3.7$ owing to the spatial fluctuation of the initial condition. That is, the transition from the uniform state is subcritical and the solitary wave state is stable even for $g<g_c$. 
For $g<3.35$, an almost uniform state was maintained at least until $t<3\times 10^6$ in our numerical simulation. Figure 1(b) shows the transition line between the almost uniform state and the solitary wave state in a parameter space of $g$ and $\beta$ obtained by the numerical simulation until $t=3\times 10^6$ starting from the random initial condition. Figure 1(c) shows a solitary wave in the mean-field model Eq.~(2) at $g=3.7$, $\beta=0.05$ and $L=500$. The solitary wave state is steadily moving in the right direction in this deterministic equation. 
The solitary wave state is maintained for $g>2.66$ at $\beta=0.05$ and $L=500$ in Eq.~(2).  The solitary wave state by Eq.~(2) is a good approximation to the steadily propagating solitary wave state in the Ising-type Vicsek model. 
\begin{figure}[t]
\begin{center}
\includegraphics[height=4.5cm]{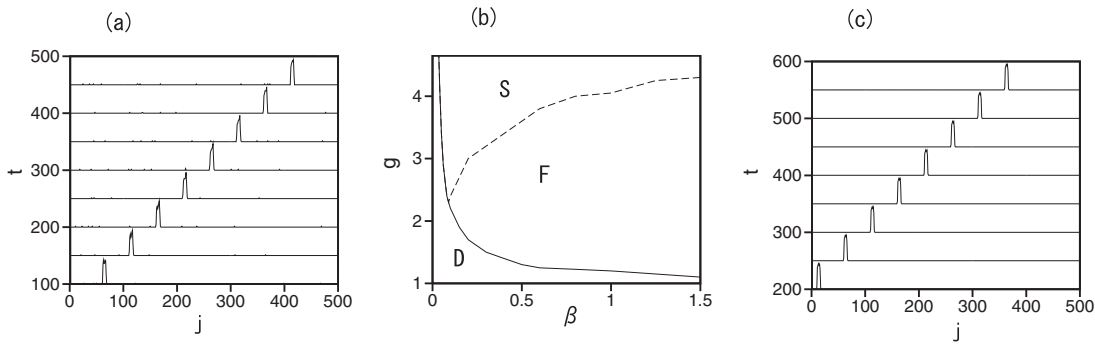}
\end{center}
\caption{(a) Time evolution of $n_{j,t}^{+}$ at $g=3.7$ and $\beta=0.05$. (b) Transition line between the almost uniform disordered state "D" and the solitary wave state "S" and the flipping solitary wave state "F" in a parameter space of $g$ and $\beta$. (c) Time evolution of $n_{j,t}^{+}$ in the mean-field model Eq.~(2) at $g=3.7$, $\beta=0.05$ and $L=500$.}
\label{f1}
\end{figure}

For $\beta>0.1$, the solitary wave exhibits a flip motion, that is, the direction of motion changes intermittently. We have investigated the parameter region where the flip motion is observed using the random initial condition. The flip motion appears in a region shown by "F" in Fig.~1(b). Note that the flip motion is  not observed in the mean-field model of Eq.~(2). That is, the fluctuations in the stochastic time evolution are important for the formation of flip motion. Figures 2(a) shows a flip motion of solitary wave at $\beta=0.15$ and $g=2.3$.  
There is one dominant solitary wave in most of the time, however, solitary waves of small amplitude appear randomly owing to the fluctuations in the stochastic time evolution. There are two solitary waves for $t<250$ in Fig.~2(a). The main solitary wave is the left-traveling one. 
Two solitary waves collide at $t\sim 250$  and a right-traveling solitary wave with large amplitude appears. The moving direction of the main solitary wave changes from left to right by the collision. The transition of moving direction of the main solitary wave occurs intermittently. In this paper, we call this phenomenon flip motion of solitary waves. Figure 2(b) shows the time evolution of $R_t=\sum_{j=1}^Lm_{j,t}/N$, which represents the order parameter for moving direction. Intermittent transitions of moving direction are clearly observed. As $g$ is increased, the transition of moving direction is suppressed and a steadily moving solitary wave appears. The parameter region of the steadily moving solitary wave state is expressed with "S" in Fig.~1(b). The phase diagram including both  the steadily moving solitary wave and the flipping solitary wave was not obtained in the model of Solon and Tailleur. 
Figure 2(c) shows a steadily moving solitary wave at $\beta=0.11$ and $g=2.3$.    
\begin{figure}[h]
\begin{center}
\includegraphics[height=5.cm]{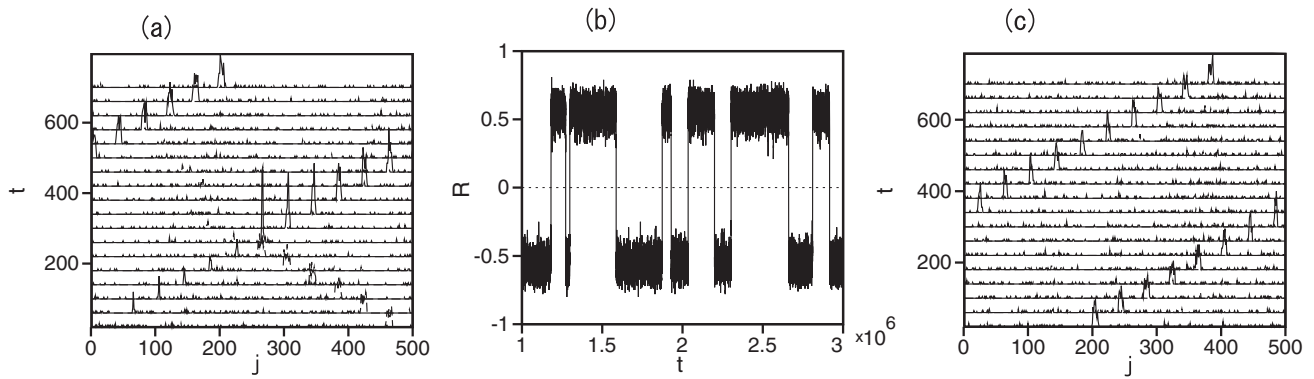}
\end{center}
\caption{(a)  Flip motion of solitary waves at $\beta=0.15$ and $g=2.3$. (b) Time evolution of $R_t=\sum_{j=1}^Lm_{j,t}/N$ at $\beta=0.15$ and $g=2.3$. (c) Right-traveling solitary wave at $\beta=0.11$ and $g=2.3$. }
\label{f2}
\end{figure}
\section{Flip motion of solitary waves}
In this section, we study the flip motion more in detail in the simplest case of Eq.~(4). The system size is set to $L=500$ and the total particle number is $N=5000$. Figure 3(a) shows the time evolution of $R_t$ at $g=2.5$. The direction of solitary wave changes intermittently. $R_t$ takes almost $+1$ or -1 at $g=2.5$, and the transition between the two states is very rapid. 
Figure 3(b) shows the time evolutions of the peak amplitudes of $n_{j,t}^{+}$ (solid green line) and $n_{j,t}^{-}$ (dashed blue line). The peak amplitude takes a maximum value just after the transition of moving direction and decreases gradually. When the peak amplitude decreases and the width of the solitary wave extends sufficiently, the transition of moving direction occurs. Figure 3(c) shows the time evolution of the profiles of $n_{j,t}^{+}$ and $n_{j,t}^{-}$ near $t=84000$. There is a left-traveling broad solitary wave of relatively small amplitude for $t<84010$.  Almost all particles are included in this left moving solitary wave. That is, the majority is the leftist. A right-traveling narrow solitary wave of  small amplitude appears in front of the left-traveling solitary wave at $t=84002$. The left-traveling solitary wave collides with the right-traveling solitary wave of small amplitude. Near the peak position of the right-traveling solitary wave, the right-traveling solitary wave has a larger amplitude than that of the left-traveling solitary wave in the front region.  Near the peak position,  left-moving particles composed of the left-traveling solitary wave turn to right with high probability $p_{+}=e^{g}/(e^{g}+e^{-g})$. Then, the peak amplitude of the right-traveling solitary wave increases rapidly owing to the turning of the left-moving particles. The left-traveling solitary wave is sequentially swallowed by the right-traveling solitary wave. The width of the right-traveling solitary wave is maintained to be narrow during the turning process.  After the right-traveling solitary wave passes over the left-traveling one, the left-traveling solitary wave almost disappears. As a result, the right-traveling narrow solitary wave with very large amplitude appears. Figuratively speaking, a moving minority appears in front of the majority. The minority swallows the majority from the frontal end, turning it into an ally sequentially. After the turning process, the minority changes to majority. However, the peak amplitude of the survived right-traveling solitary wave decreases gradually and the solitary wave becomes broad after a long time. Then, another seed of left-traveling solitary wave appears and overthrows the right-traveling solitary wave vice versa. Thus, the transitions between the rightist and leftist regimes repeat infinitely as shown in Fig.~3(a).   The creation process of a sharp solitary wave by the collision of inversely-traveling waves is important for the formation of a solitary wave in a population of self-propelled particles, regardless of the presence or absence of the flip motion. 
  
\begin{figure}[h]
\begin{center}
\includegraphics[height=4.5cm]{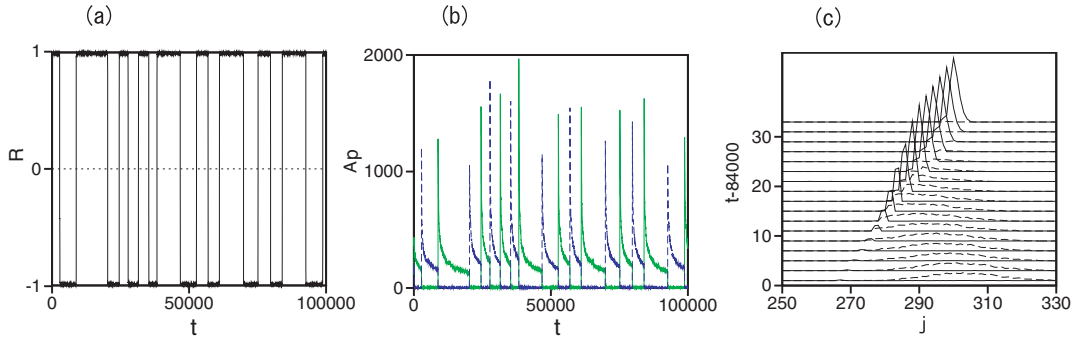}
\end{center}
\caption{(a) Time evolution of $R_t$ at $g=2.5$. (b) Time evolutions of the peak amplitudes of $n_{j,t}^{+}$ (solid green line) and $n_{j,t}^{-}$ (dashed blue line). (c) Time evolutions of the profiles of $n_{j,t}^{+}$ and $n_{j,t}^{-}$ near $t=84000$. }
\label{f3}
\end{figure}

As shown in Fig.~3(b), the peak amplitude decreases after the transition of moving direction.  
Figure 4(a) shows the time evolution of the peak amplitude $A_p$ at $g=2.5$ in a semi-logarithmic plot after $t=t_c=84030$. The dashed line is $1650\times e^{-0.0045(t-t_c)}$ and the dotted line is $1250/(t-t_c)^{1/2}$. That is, the peak amplitude initially decays exponentially, but it decays in a power law for large $t-t_c$. The long-time decay of $1/t^{1/2}$ is similar to that in the diffusion process. 
The average reversal time $T$ can be evaluated as an average time between successive two time steps satisfying $R_t=0$. 
Figure 4(b) shows the average reversal time $T$ as a function of $g$. The dashed line is $37.5e^{2g}$. The average reversal time increases exponentially with $g$ for $g<2.6$. 

For $\tilde{m}_{j,t}>0$, Eq.~(2) is approximated at
\begin{eqnarray}
n_{j+1,t+1}^{+}&=&p_{+}n_{j,t},\nonumber\\
n_{j-1,t+1}^{-}&=&(1-p_+)n_{j,t},
\end{eqnarray}
where $p_{+}=e^g/(e^g+e^{-g})$. We have performed numerical simulation of Eq.~(8) using an initial condition $n_{L/2,0}^{+}=n_{L/2+1,0}^{+}=1650$ and $n_{j,0}^{+}=0$ for the other $j$'s and $n_{j,0}^{-}=0$ for all $j$'s. 
Figure 4(c) shows the peak amplitude $A_p$ of $n_{j,t}^{+}$ at $g=2.5$. The dashed line is $1650\times e^{-\alpha t}$ with $\alpha=0.0067$ and the dotted line is $2000/t^{1/2}$. That is, even in this deterministic mean-field model, 
the peak amplitude decays initially exponentially and decays in a power law for large $t$.  Figure 4(d) shows the decay time $T$ as a function of $g$ in this deterministic model. Here, the decay time is evaluated at a time where the peak amplitude decays to $A_p(T)=107$. The dashed line is $37.5e^{2g}$, which is very close to the numerical result shown in Fig.~4(b), although $A_p(T)=107$ is chosen as a fitting parameter.  Since Eq.~(8) is a linear equation, the decay time is independent of the initial amplitude $A_p(0)$ and the decay time approximated at $37.5e^{2g}$ comes from the decay ratio $A_p(T)/A_p(0)=107/1650\sim 0.065$. The decay time is well approximated with the average reversal time shown in Fig.~4(b) for the Ising-type Vicsek model, that is, the decay of the peak amplitude can be qualitatively understood by this mean-field approximation. However, from the time sequence shown in Fig.~3(b), the ratio of the peak amplitude of $n_{j,t}^{+}$ just after the transition and just before the next transition can be evaluated as 0.12, which is rather larger than the value 0.065 for the deterministic mean-field model. We do not understand the reason yet completely, but some effect of fluctuations neglected in the mean-field model might be important.  
If the minority particles are neglected, Eq.~(8) is further approximated as 
\[
n_{j+1,t+1}^{+}=p_{+}n_{j,t}^{+}.\]
Then, $R_t$ is approximated as $R_t\propto R_{t_c}e^{-\alpha(t-t_c)}$ where $\alpha=-\log\{e^g/(e^g+e^{-g})\}\sim 0.0067$ for $g=2.5$. This exponent is consistent with numerical simulation shown in Fig.~4(c), however, it is larger than the exponent 0.0045 for the dashed line in Fig.~4(a).  
\begin{figure}[h]
\begin{center}
\includegraphics[height=4.cm]{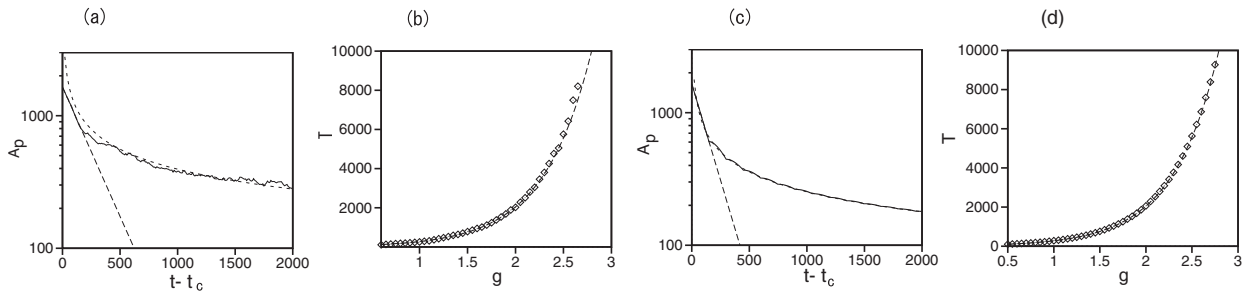}
\end{center}
\caption{(a) Time evolution of the peak amplitude $A_p$ in a semi-logarithmic plot after $t_c=84030$ in the Ising-type Vicsek model. The dashed line is $1650\times e^{-0.0045(t-t_c)}$ and the dotted line is $1250/(t-t_c)^{1/2}$. (b) Average reversal time $T$ as a function of $g$. The dashed line is $37.5e^{2g}$. (c)
 Time evolution of the peak amplitude $A_p$ in the deterministic mean-field model Eq.~(8) at $g=2.5$. The dashed line is $1650\times e^{-\alpha t}$ with $\alpha=0.0067$ and the dotted line is $2000/t^{1/2}$. (d) Decay time $T$  from $A_p(0)=1650$ to $A_p(T)=107$ as a function of $g$ in this deterministic model. The dashed line is $37.5e^{2g}$.}
\label{f4}
\end{figure}
\begin{figure}[h]
\begin{center}
\includegraphics[height=4.cm]{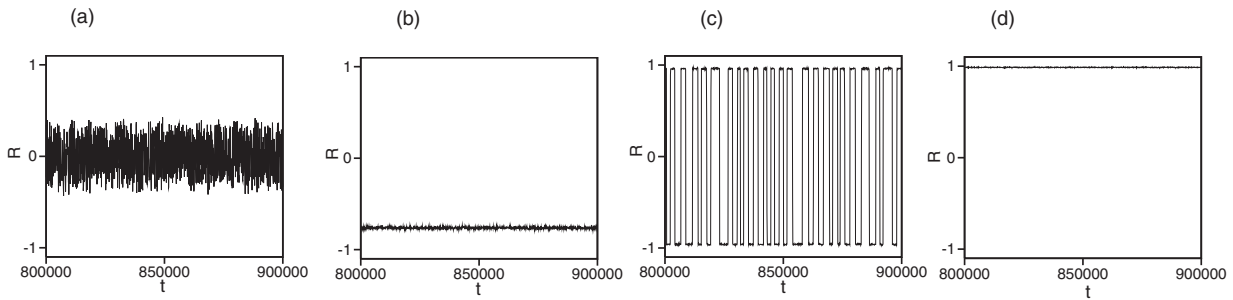}
\end{center}
\caption{Time evolutions of $R_t$ at (a) $g=0.5$, (b) $g=1.0$, (c) $g=2$, and  (d) $g=2.5$ for $N=5000$ and $L=300$. }
\label{f5}
\end{figure}
\begin{figure}[h]
\begin{center}
\includegraphics[height=4.cm]{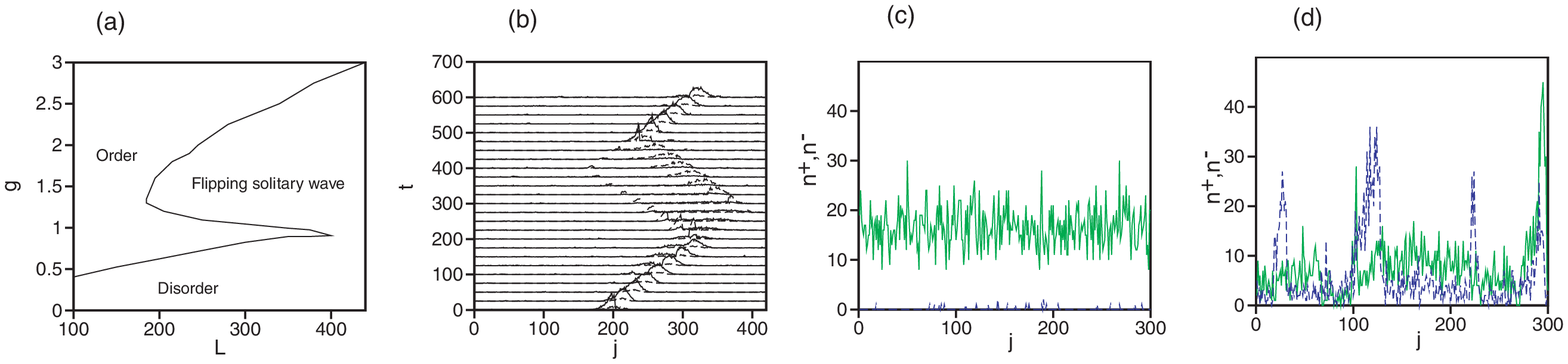}
\end{center}
\caption{(a) Phase diagram in a parameter space of $L$ and $g$ for $N=5000$. (b) Time evolution of the profiles $n_{j,t}^{+}$ and $n_{j,t}^{-}$ at $g=0.8$ for $L=420$. (c) Snapshot profile $n_{j,t}^{+}$ (solid green line) and $n_{j,t}^{-}$ (dashed blue line) in an ordered state for $g=2.5$ for $L=300$. (d) Snapshot profile $n_{j,t}^{+}$ and $n_{j,t}^{-}$ in an intermediate state at $g=0.5$ and $L=420$.}
\label{f6}
\end{figure}

Flip motions of solitary waves are observed in a wide range of parameters. 
We have changed two parameters $L$ and $g$ for $N=5000$ to investigate the parameter range for flip motion. Figures 5(a)-(d) show the time evolutions of $R_t$ at (a) $g=0.5$, (b) $g=1.0$, (c) $g=2$, and  (d) $g=2.5$ for $L=300$. The disordered state appears at $g=0.5$, and the ordered state appears at $g=1$ and 2.5. In these ordered states, the symmetry breaking characterized by $|R_t|>0$ occurs, however, there is no localized structure or an almost spatially-uniform state appears. Flip motion of solitary wave is observed at $g=2$. Both in the disordered state and the flipping state of solitary waves, the sign of $R_t$ changes randomly around $0$. In the disordered state, $R_t$ is fluctuating near $R_t=0$, and in the flipping state of solitary wave, $R_t$ spends most of the time near +1 or -1. In the flipping state of solitary wave, one dominant solitary wave appears and the direction changes intermittently. 

Figure 6(a) is a phase diagram in the parameter space of $L$ and $g$ for $N=5000$. Numerical simulations were performed for the random initial condition. There are three states: the disordered state, ordered state, and flipping solitary wave state.  The transition line between the disordered state and the flipping solitary wave state is unclear, because $R_t$ is fluctuating around $R=0$ in both states. Figure 6(b) shows the time evolution of the profiles $n_{j,t}^{+}$ and $n_{j,t}^{-}$ at $g=0.8$ for $L=420$, which is a randomly flipping solitary wave state. Figure 6(c) shows a snapshot of the ordered state at $g=2.5$ for $L=300$. 
The profiles of $n_{j,t}^{+}$ and $n_{j,t}^{-}$ are almost uniform. $n_{j,t}^{+}$ is fluctuating around $16$ and $n_{j,t}^{-}$ is almost 0. Figure 6(d) shows a snapshot of an intermediate state between the disordered state and the flipping solitary wave state at $g=0.5$ and $L=420$. Several small peaks and random background fluctuations are observed both for $n_{j,t}^{+}$ and $n_{j,t}^{-}$. 
\section{Flip motion in the two-dimensional Ising-type Vicsek model}
The Ising-type Vicsek model can be generalized to a two-dimensional system. In the two-dimensional model, particles are randomly set on a rectangular lattice of $L_x\times L_y$, and the velocity takes one of two values $+1$ and $-1$ randomly as an initial condition. Periodic boundary conditions are assumed in the $x$ direction (horizontal direction).  Only the case of Eq.~(4) ($\beta=\infty$) is considered in this section. For the interaction, $\tilde{m}_{j,k,t}$ at the $(j,k)$ site is defined as $\tilde{m}_{j,k,t}=m_{j,k,t}+q(m_{j+1,k,t}+m_{j-1,k,t}+m_{j,k+1,t}+m_{j,k-1,t})$ where $m_{j,k,t}=n_{j,k,t}^{+}-n_{j,k,t}^{-}$ and $q=1/2$.  No-flux boundary conditions such as $n_{j,L_y+1,t}^{+}=n_{j,L_y,t}^{+}$ and $n_{j,0,t}^{-}=n_{j,1,t}^{-}$ are assumed in the $y$ direction for the calculation of $\tilde{m}_{j,k,t}$. If $\tilde{m}_{j,k,t}>0$,  the velocity of a particle at the $(j,k)$ site takes $+1$ at the next time step $t+1$ with probability $p_{+}=e^g/(e^g+e^{-g})$, and  $-1$ with probability $p_{-}=e^{-g}/(e^g+e^{-g})$ similarly to the one-dimensional model. 

We have performed numerical simulation in a rectangle of system size $L_x\times L_y=500\times 50$. In each row, there are $N=5000$ particles, therefore, the total particle number is $5000\times 50=250000$. Particles do not move to different rows, because the velocity takes only $+1$ or $-1$.   Figure 7(a) shows the time evolution of $R_t$ at $g=1.5$. The sign of $R_t$ changes randomly. Figure 7(b) shows a snapshot of $n_{j,k,t}^{+}$ (denoted by solid green line) and $n_{j,k,t}^{-}$ (denoted by dashed blue line) at $n=1194100$. There is one dominant solitary wave in most of the time in each row. The transitions of traveling direction occur intermittently similarly to the one-dimensional model. 
The solitary waves make a cluster among the neighboring sites in the $y$ direction and move together in the $x$ direction. There are eight to nine clusters of two-dimensional solitary waves in this snapshot, however, the clustering changes with time. As $g$ increases, the number of clusters decreases. 
\begin{figure}[h]
\begin{center}
\includegraphics[height=4.5cm]{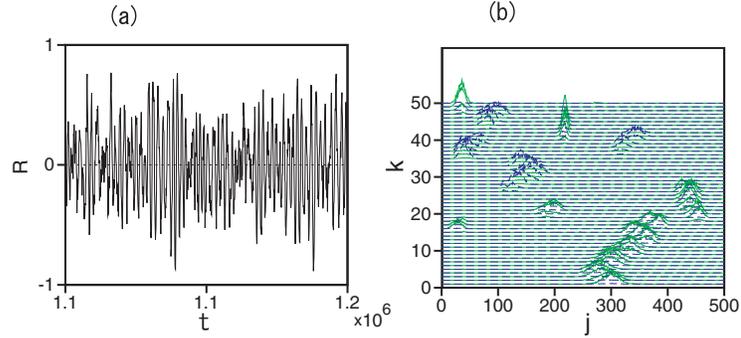}
\end{center}
\caption{(a) Time evolution of $R_t$ at $g=1.5$ in the two-dimensional Ising-type Vicsek model. (b) Snapshot of $n_{j,k,t}^{+}$ (solid green line) and $n_{j,k,t}^{-}$ (dashed blue line) at $t=1194100$.}
\label{f7}
\end{figure}

Figure 8(a) shows the time evolution of $R_t$ at $g=2.8$. The transition is not so sharp compared to Fig.~3(a). This is because the transition does not occur simultaneously in the $y$ direction as shown later in Fig.~9. Figure 8(b) shows the time evolution of the peak amplitudes of $n_{j,k,t}^{+}$ and $n_{j,k,t}^{-}$. Similarly to the one-dimensional model, a sharp solitary wave appears just after the transition and then the peak amplitude decays gradually with time. When the peak amplitude is sufficiently small, an inversely-traveling small two-dimensional solitary wave appears and the minor solitary wave overthrows the major solitary wave in the same way as the one-dimensional model. However, the velocity-turning process occurs sequentially in the $y$ direction. Figures 9 shows four snapshots in the turning process at (a) $t=1180000$, (b) 1181050, (c) 1181800, and (d) 1190000 for $g=2.8$. A band-like soliton moves in the left at $t=1180000$.
A very sharp right-traveling solitary wave appears and the regime change occurs locally near $k=40$ at $t=1181050$.  The turning process extends in the $y$-direction or the sharp right-traveling solitary wave extends to the neighboring $k$ sites as shown in Fig.~9(c). It takes some time for the velocity-turning process to complete in the $y$ direction. 
Finally, another bandlike soliton appears. Figure 9(d) shows a right-traveling  solitary wave at $t=1190000$.    

\begin{figure}[h]
\begin{center}
\includegraphics[height=4.5cm]{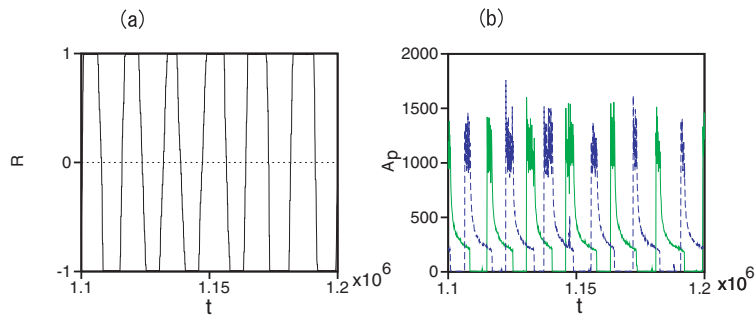}
\end{center}
\caption{(a) Time evolution of $R_t$ at $g=2.8$ in the two-dimensional Ising-type Vicsek model. (b) Time evolution of the maximum values of $n_{j,k,t}^{+}$ (solid green line)  and $n_{j,k,t}^{-}$ (dashed blue line) at $g=2.8$.}
\label{f8}
\end{figure}
\begin{figure}[h]
\begin{center}
\includegraphics[height=4.cm]{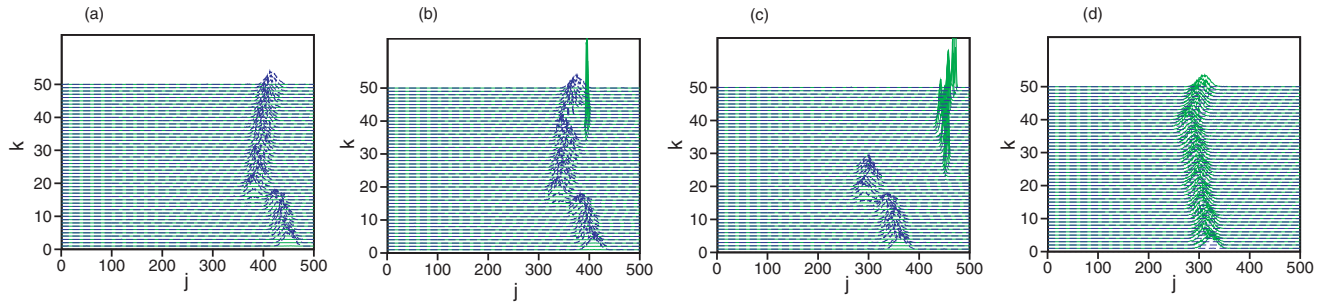}
\end{center}
\caption{Four snapshots of $n_{j,k,t}^{+}$ (solid green line) and $n_{j,k,t}^{-}$ (dashed blue line) at (a) $t=1180000$, (b) 1181050, (c) 1181800, and (d) 1190000 for $g=2.8$.}
\label{f9}
\end{figure}
\section{Summary and discussion}
We have proposed the Ising-type Vicsek model for the collective motion and sudden direction change of self-propelled particles. Particles move on a lattice with velocity $+1$ or $-1$.  
Our model is similar to the active Ising model by Solon and Tailleur in that particles move in the right or left direction on the lattice, however, 
the detailed updating rules are different. The probability of the velocity of a particle on a lattice site at the next step is determined by the number difference between the right- and left- moving particles at the lattice site and its nearest-neighboring sites. We have found that flip motions of the main solitary wave occur intermittently. An inversely-moving solitary wave with small amplitude (minority) appears in front of the main broad solitary wave. The inversely moving minority attacks the majority and particles in the main solitary wave change their direction sequentially. As a result of the surge of turning process, the minority wins the majority and the inversely traveling soliton appears. The peak amplitude of the solitary wave is very sharp just after the regime change, however, it decays gradually and the solitary wave becomes broad. After the broadening process, another regime change occurs again. We have proposed a simple mean-field equation, which can explain qualitatively the gradual decay of the peak amplitude of the solitary wave. We have further generalized the one-dimensional model to a two-dimensional model, and found similar flip motions even in the two-dimensional model, which was not observed in the active Ising model. 

The rapid growth of a solitary wave by swallowing inversely-traveling solitary waves is an important process of the formation of a solitary wave structure regardless of the flipping solitary wave state or steadily-moving solitary wave state in the Ising type models. The number of solitary waves decreases and one dominant solitary wave appears after the initial transient time by the swallowing process. On the other hand, the microphase separation occurs and many band solitons appear in the original Vicsek model~\cite{Solon3}. 

We have shown only the case of the nearest-neighbor interaction in this paper, however, similar flip motions of solitary waves are observed even in a model with both the nearest-neighbor and next-nearest-neighbor couplings.  
We have also confirmed that similar flipping solitary wave states appear even when the probability Eq.~(4) is replaced by $p_{j,t+1}^{\pm}=e^{\pm g\tilde{m}_{j,t}/\tilde{n}_{j,t}}/(e^{g\tilde{m}_{j,t}/\tilde{n}_{j,t}}+e^{-g\tilde{m}_{j,t}/\tilde{n}_{j,t}})$ where $\tilde{n}_{j,t}=n_{j,t}+q(n_{j-1,t}+n_{j+1,t})$.  Owing to the division by $\tilde{n}_{j,t}$, inversely moving particles are generated with a finite probability even if the particles moving in the same direction strongly gather together and form a solitary wave. The finite probability for the nucleation of a seed of the inversely traveling solitary wave is important for the formation of flip motions. 

It is left to future study to perform some more mathematical analysis to our simple model and generalize the model to reproduce the sudden cooperative change of moving direction in realistic self-propelled particles such as the flock of starling. 

\section*{Acknowledgment}

This work was supported by a Grant-in-Aid for Scientific Research (No. 18K03462) from the Ministry of Education, Culture, Sports, Science and Technology of Japan.


\begin{thebibliography}{99}
\bibitem{Vicsek} T.~Vicsek, A.~Czir\'ok, E.~Ben-Jacob, I.~Cohen, and O.~Shochet, Phys. Rev. Lett. {\bf 75}, 1226 (1995).
\bibitem{Vicsek2} T.~Vicsek and A.~Zafeiris, Phys. Rep. {\bf 517}, 71 (2012).
\bibitem{Marchetti} M.~C.~Marchetti, J.~F.~Joanny, S.~Ramaswamy, T.~B.~Liverpool, J.~Prost, M.~Rao, and R.~A.~Simha, Rev. Mod. Phys. {\bf 85}, 1143 (2013).
\bibitem{Sokolov} A.~Sokolov, I.~S.Aranson, J.~O.~Kessler, R.~E.~Goldstein, Phys. Rev. Lett. {\bf 98}, 158102 (2007). 
\bibitem{Scaller} V.~Schaller, C.~Weber, C.~Semmrich, E.~Frey, and A.~R.~Bausch, Nature {\bf 467}, 73 (2010).
\bibitem{Katz} Y.~Katz, K.~Tunstr{\o}m, C.~C.~Ioannou, C.~Huepe, and I.~D.~Couzin, Proc. Nat. Acad. Sci. {\bf 108}, 18720 (2011).
\bibitem{Thutupalli} S.~Thutupalli, R.~Seemann, and S.~Herminghaus, New J. Phys. {\bf 13}, 073021 (2011).
\bibitem{Sumino} Y.~Sumino, K.~H.~Nagai, Y.~Shitaka, D.~Tanaka, K.~Yoshikawa, H.~Chat\'e, and K.~Oiwa, Nature {\bf 483}, 448 (2012).
\bibitem{gre} G.~Gr\'egoire and H.~Chat\'e, Phys. Rev. Lett. {\bf 92}, 025702 (2004).
\bibitem{Chate} H.~Chat\'e, F.~Ginelli, G.~Gr\'egoire, F.~Peruani, and F.~Raynaud, Eur. Phys. J. B {\bf 64}, 451 (2008). 
\bibitem{Bertin} E.~Bertin, M.~Droz, and G.~Gr\'egoire, Phys. Rev. E {\bf 74}, 022101 (2006).
\bibitem{Mishra} S.~Mishra, A.~Baskaran, and M.~C.~Marchetti, Phys. Rev. E {\bf 81}, 061916 (2010).
\bibitem{Gopinath} A.~Gopinath, M.~F.~Hagan, M.~C.~Marchetti, and A.~Baskaran, Phys. Rev. E {\bf 85}, 061903 (2012).
\bibitem{Bricard} A.~Bricard, J.~B.~Caussin, N.~Desreumaux, O.~Dauchot, and D.~Bartolo, Nature {\bf 503}, 95 (2013). 
\bibitem{Ihle} T.~Ihle, Phys. Rev. E {\bf 88}, 040303(R) (2013).
\bibitem{Caussin} J.~B.~Caussin, A.~Solon, A.~Peshkov, H.~Chat\'e, T.~Dauxois, J.~Tailleur, V.~Vitelli, and D.~Bartolo, Phys. Rev. Lett. {\bf 112}, 148102 (2014).\bibitem{Sakaguchi} H.~Sakaguchi and K.~Ishibashi, J. Phys. Soc. Jpn. {\bf 86}, 114003 (2017).
\bibitem{Sakaguchi2} H.~Sakaguchi and K.~Ishibashi, J. Phys. Soc. Jpn. {\bf 87}, 064001 (2018).
\bibitem{Solon} A.~P.~Solon and J.~Tailleur, Phys. Rev. Lett. {\bf 111}, 078101 (2013).
\bibitem{Solon2} A.~P.~Solon and J.~Tailleur, Phys. Rev. E {\bf 92}, 042119 (2015). 
\bibitem{Solon3} A.~P.~Solon, H.~Chat\'e, and J.~Tailleur, Phys. Rev. Lett. {\bf 114}, 068101 (2015). 
\end{thebibliography}
\end{document}